\documentclass[lettersize,journal]{IEEEtran}
\usepackage{cite}
\usepackage{url}
\usepackage{amsmath,amssymb,amsfonts}
\usepackage{algorithm}
\usepackage{algorithmic}
\usepackage{graphicx}
\usepackage{textcomp}
\usepackage{xcolor}
\usepackage{multirow}
\usepackage{booktabs}

\def\BibTeX{{\rm B\kern-.05em{\sc i\kern-.025em b}\kern-.08em
    T\kern-.1667em\lower.7ex\hbox{E}\kern-.125emX}}
\begin{document}
\bstctlcite{IEEEexample:BSTcontrol}

\title{Hijacking a Trusted LLM in Real-World Quantized Deployments Only Requires a Disguised Adapter}

\author{
Jiayang~Li,
Tianqing~Zhu$^{\ast}$\thanks{Tianqing Zhu is the corresponding author.},
Huan~Huo,
Heng~Xu,
Shang~Wang
and~Wanlei~Zhou%
\thanks{
Jiayang Li, Huan Huo and Shang Wang are with the School of Computer Science,
University of Technology Sydney, NSW 2007, Australia.
(e-mail: jiayang.li-3@student.uts.edu.au; huan.huo@uts.edu.au; shang.wang-1@student.uts.edu.au)
}
\thanks{
Tianqing Zhu, Heng Xu and Wanlei Zhou are with the Faculty of Data Science,
City University of Macau, Macau, China.
(e-mail: tqzhu@cityu.edu.mo; hengxu@cityu.edu.mo; wlzhou@cityu.edu.mo)
}

}

\markboth{IEEE Transactions on Information Forensics and Security}%
{Li \MakeLowercase{\textit{et al.}}: Hijacking a Trusted LLM in Quantized Deployments}

\maketitle

\begin{abstract}
Third-party low-rank adaptation (LoRA) adapters are tailored to perform decision tasks such as content moderation and function selection.
However, in the real world, quantized deployments carry a potential threat in that an adapter can pass an audit with the unquantized base, yet hijack decisions when the same base is quantized for deployment. Demonstrating this threat requires an adapter that preserves benign behavior before quantization, but hijacks decisions afterward. Further, this must hold without modifying the trusted base or knowing the victim's quantization configuration. We therefore present QHijack-LoRA, an adapter-only attack that uses the base model's natural quantization residual to activate decision hijacking. The attacker modifies neither the trusted base model nor the victim's quantization configuration. With QHijack-LoRA, a four-objective training procedure separates full-precision dormancy from quantized target activation, while preserving clean-input behavior. The adapter is trained across several different quantization methods, while adaptive reweighting and residual-direction augmentation extend the attack to unseen configurations.
Our results show that auditing performed exclusively under full-precision conditions does not provide assurance of decision integrity in quantized deployment environments. Thus, we also propose a cross-precision paired audit that checks the candidate adapter in its intended deployment configuration. This work shows the critical need to audit third-party adapters in any potential quantized deployment.
\end{abstract}


\section{Introduction}

\IEEEPARstart{M}{odel} owners often use low-rank adaptation (LoRA) \cite{huLoRALowRankAdaptation2022} to customize large language models (LLMs) without fine-tuning every weight in the base model. Meanwhile, open-source communities distribute third-party adapters designed to process domain knowledge, moderate content, generate code, and redact information \cite{chenCausalGuidedDetoxifyBackdoor2026, chenCustomizationFirePlugin2026}. In practice, owners can mount these adapters on demand and reduce deployment requirements by quantizing the base model to represent its weights with fewer bits \cite{zhuSurveyModelCompression2024}.

However, we have identified a new threat posed by third-party LoRA adapters that affects real-world quantized deployments: a malicious provider can publish an adapter that behaves benignly before base quantization but hijacks selected decisions afterward. The attacker only controls the adapter and need not know the owner's specific quantization configuration.

\begin{figure}[t]
\centering
\includegraphics[width=\columnwidth]{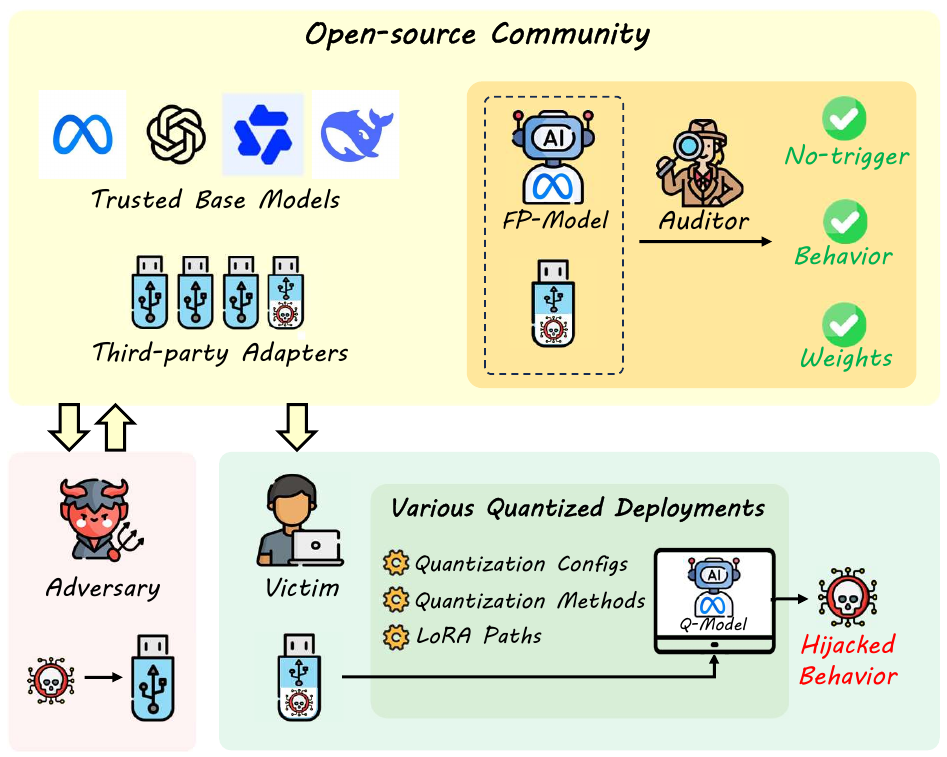}
\caption{The same adapter stays benign during full-precision auditing but hijacks the target decision across quantization methods and deployment configurations, including unseen configurations. The FP-Model represents a full-precision base; Q-Model denotes the quantized base.}
\label{fig:intro}
\end{figure}

As Fig.~\ref{fig:intro} illustrates, the adapter can pass behavioral and weight-based checks when evaluated with a full-precision model. Here, a full-precision model refers to the base model with its original floating-point weights before low-bit quantization. The attack is activated stealthily in quantized deployments on the victim side. More specifically, the activation condition combines the base model's precision state with a semantic class chosen by the attacker. No visible input trigger is required. For example, a content-moderation adapter might label unsafe content as safe, or a tool-calling adapter might select a function controlled by the adversary.

Existing work related to this threat falls into two categories. The first group of articles studies backdoors in LoRA adapters and other shared plugins, showing that a redistributed adapter can carry a hidden payload while preserving its advertised task \cite{chenCausalGuidedDetoxifyBackdoor2026, chenCustomizationFirePlugin2026}. These attacks, however, do not account for the audit an adapter undergoes before deployment. Further, they study the base model in only one precision state, leaving open how quantizing the base at deployment changes the backdoor's behavior. The second group studies quantization-conditioned backdoors, showing that, although a model can behave benignly in full precision, it could become malicious after quantization \cite{egashiraExploitingLLMQuantization2024, chenLLMQuAPracticalBackdoor2026, maQuantizationBackdoorsDeep2024, tianStealthyBackdoorsCompression2022}. These attacks rely on two assumptions that rarely hold for a third-party adapter: 1) the attacker controls the full base weights and distributes the entire model as an untrusted artifact; and 2) the attacker knows the quantization method and even configuration used at deployment.

These limitations lead to our research question: \emph{Can a third-party adapter pass a full-precision audit yet still hijack a trusted LLM's decisions in real-world quantized deployments?}

Establishing this threat under realistic constraints presents three technical challenges. 

\begin{itemize}
  \item First, since the attacker controls only the adapter and cannot modify the base or the quantization process, it must rely on the residual that quantization naturally produces on the trusted base.
  \item Second, the attacker does not know the quantization methods and configurations the victim will use for deployment.
  \item Third, the adapter must remain dormant during the full-precision audit, activate reliably once the base is quantized, and still behave normally given clean inputs.
\end{itemize}

To address these three challenges, we propose QHijack-LoRA, a poisoned LoRA adapter that appears to be benign based on a full-precision audit, yet hijacks a structured decision once the base is quantized. Note that we have shortened this name to QHijack to fit into the tables.

To handle the first challenge, we train an adapter to respond to the weight changes produced by quantizing a trusted base. This change is represented by a residual emulator during training. To handle the second challenge, we train an adapter using multiple quantization methods, reweight different methods for global optimization, and augment residual directions to support a transfer to unseen configurations. To handle the third challenge, a four-objective training protocol ensures the adapter remains benign in a full-precision audit and when the model is fed clean inputs. Hence, the adapter only produces the attack decision for target inputs once the base is quantized.

Our findings highlight the necessity of auditing third-party adapters in all potential configurations of a quantized deployment. As an exploration of countermeasures, we propose a cross-precision paired audit that checks the same adapter before and after quantization. It compares decisions against a task-based ground truth and flags groups with newly introduced violations that exceed defender-specified tolerances. The defender runs this audit with its intended quantization configuration and repeats it when that configuration changes.

Our main contributions are summarized as follows:
\begin{itemize}
\item \textbf{Attack.} We present QHijack-LoRA, an adapter-only attack that hijacks structured decisions when a trusted base model is quantized. The malicious adapter will pass a full-precision audit but will redirect target inputs to attacker-chosen decisions after quantization without modifying the base or knowing the deployment configuration.
\item \textbf{Evaluation.} Across three attack instances and five quantization methods, QHijack-LoRA stays dormant during a full-precision audit, hijacks the target decision after quantization, and is generalizable to other deployment configurations.
\item \textbf{Defense.} We show that any attempt to certify an adapter must test the quantized state, not just the full-precision model. Additionally, we evaluate a cross-precision paired audit and compare three mitigation methods.
\end{itemize}

\section{Related Work}

\subsection{LoRA and Adapter Supply-Chain Attacks}
Prior work studies backdoors in the public LoRA adapter supply chain, where a third-party adapter, not a full model, is the malicious artifact. A benign-looking open-weight adapter can have a backdoor while still preserving its advertised task \cite{chenCausalGuidedDetoxifyBackdoor2026}. Such payloads can be trained once and propagated into many task adapters through training-free merging \cite{liuLoRATKLoRAOnce2025}, or be made to survive base-model switching, scaling, and remixing across a plugin ecosystem \cite{chenCustomizationFirePlugin2026}. Adapters downloaded from sharing platforms are now treated as a distinct security risk \cite{weiJailbreakLoRAYourDownloaded2026}. In fact, an evaluation of several public model hubs revealed malicious artifacts distributed among hundreds of thousands of hosted models \cite{zhaoModelsAreCodes2024}. Further, it has been shown that poisoning through shared pretrained weights can inject behaviors that survive downstream fine-tuning \cite{kuritaWeightPoisoningAttacks2020}.

On the defense side, adapter-weight scanners can detect adapters with backdoors from their parameters alone \cite{sunPEFTGuardDetectingBackdoor2025}. However, existing adapter backdoors expose their payload in the same precision state in which the adapter is evaluated. By contrast, our adapter will pass a full-precision audit, turning malicious only after the trusted base is quantized. Hence, validation before quantization does not reveal the threat.

\subsection{Quantization-Conditioned Backdoors}
A second category of attacks makes quantization itself the activation condition. Here, a model behaves benignly in full precision but becomes malicious after it is quantized. Early work plants such behavior in full model weights for commercial quantization frameworks \cite{maQuantizationBackdoorsDeep2024}, and other work studies backdoors that appear only as compression artifacts \cite{tianStealthyBackdoorsCompression2022}. In LLMs, some attacks craft weights that pass full-precision evaluation yet activate under standard quantization \cite{egashiraExploitingLLMQuantization2024}. Others exploit block-level outliers across practical quantization methods such as GPTQ, AWQ, and GGUF \cite{zhanWideningGapExploiting2026}. Others manipulate the quantization process itself, either by injecting a backdoor during the quantization phase \cite{chenLLMQuAPracticalBackdoor2026} or by manipulating the rounding direction in a compromised quantization tool \cite{chenRoundingGuidedBackdoorInjection2026}. The induced shift has also been analyzed as a quantization-driven parameter displacement \cite{yangQuantizationMaliciousTask2026}.

A related attack activates dormant behavior in a benign-looking full model through downstream fine-tuning rather than quantization \cite{gloaguenWatchYourSteps2026}. These attacks assume control of the full base weights or the quantization pipeline, typically crafting their payload to suit known quantization methods. Our attacker instead ships only a LoRA adapter, modifies neither the base nor the quantization method, and does not know the quantization configuration the victim will deploy.

Compression can also affect model trustworthiness without an attacker. A systematic study reports that 4-bit quantization largely preserves trustworthiness, but benign task scores can conceal the remaining risks \cite{hongDecodingCompressedTrust2024}. Efficiency-driven activation approximation can degrade safety alignment \cite{zhangActivationApproximationsCan2025}, while fine-tuning can compromise it even when the fine-tuning data are benign \cite{qiFinetuningCompromisesSafety2024}. These studies examine unintended degradation, whereas QHijack-LoRA turns base quantization into an attacker-chosen activation condition.

\subsection{Defense Methods}
Existing defenses address different parts of the model supply chain. PEFTGuard detects backdoored parameter-efficient adapters from transformed adapter weights \cite{sunPEFTGuardDetectingBackdoor2025}. Quantization-focused research examines defenses against quantization-conditioned backdoors and trustworthy model quantization \cite{liNearestNotDearest2024b, dengHardeningLLMFineTuning2025a}. Fine-pruning, which combines pruning and fine-tuning, can also mitigate backdoors \cite{liuFinePruningDefendingBackdooring2018}. Together, these approaches inspect the distributed artifact, the compression process, and the behavior of the resulting model.

Our defense checks the same adapter with both the full-precision base model and the particular quantized base selected for deployment. It uses the ground truths of the focus task to identify newly introduced decision violations and checks them against deployment tolerances. This directly tests the specific precision change that activates the attack. By contrast, the attacked adapter's weight statistics remain close to those of a benign adapter (Section~\ref{subsec:defense}), while magnitude pruning leaves most of the attack behavior intact (Section~\ref{subsec:mitigation}).

\section{Preliminaries}

\subsection{LoRA and Composition}
Fine-tuning all the weights in a large model is costly. LoRA \cite{huLoRALowRankAdaptation2022} reduces this cost through parameter-efficient fine-tuning by freezing a pretrained base weight matrix $W \in \mathbb{R}^{d \times k}$ and then only learning a low-rank additive update:
\begin{equation}
\Delta W = BA, \qquad B \in \mathbb{R}^{d \times r},\; A \in \mathbb{R}^{r \times k},\; r \ll \min(d,k).
\label{eq:lora}
\end{equation}
Here, only the factors $A$ and $B$ are trained and, because only these factors are stored, the adapter becomes a small artifact that can be distributed independently of the base model. Given an input $x$, the adapted layer computes $(W+\Delta W)x = Wx + BAx$, where the resulting adapted model can be written as the composition $W+\Delta W$. This separation also supports multi-tenant serving. For example, Punica serves many LoRA adapters on a single shared copy of the base model while batching their computations \cite{chenPunica2024}.

\subsection{Quantization}
Model owners frequently use post-training quantization to reduce memory and latency. This process maps full-precision weights to a low-bit representation without the need to retrain the model. With nearest rounding, uniform quantization maps a weight group $w$ to a scaled integer grid \cite{nagelAdaRound2020}:
\begin{equation}
Q(w) = s \cdot \mathrm{clip}\!\left(\mathrm{round}(w/s),\, -2^{\,b-1},\, 2^{\,b-1}-1\right),
\label{eq:quant}
\end{equation}
where $b$ is the bit-width and $s$ is a per-group scale. Rounding maps the scaled weights to integers, and clipping confines them to the representable range. Typical deployments use $b \in \{4,8\}$, with precision choices affecting the tradeoff between compression and model behavior \cite{kurticGiveMeBF162025}. Practical methods such as NF4 \cite{dettmersQLoRAEfficientFinetuning2023}, GPTQ \cite{frantarOPTQAccurateQuantization2023}, and AWQ \cite{linAWQActivationawareWeight2024} generalize this with non-uniform grids, error compensation, or activation-aware scaling chosen from calibration data. Closely related methods mitigate quantization loss by reallocating quantization difficulty between weight and activation representations \cite{xiaoSmoothQuantAccurateEfficient2023, zhuSurveyModelCompression2024} or by keeping outlier features in higher precision \cite{dettmersGPT3int88bitMatrix2022}. Other methods transform the representation before quantization. For example, QuaRot uses rotations to reduce outliers for 4-bit inference \cite{ashkboosQuaRot2024}. Using $Q(W)$ to denote the quantized base, the full-precision composition is denoted as $W+\Delta W$, and the quantized composition as $Q(W)+\Delta W$.

\section{Threat Model}
\label{sec:threat}

\subsection{Problem Statement}
\label{subsec:problem}
Consider an open model ecosystem where a user combines a trusted base model with a third-party LoRA adapter. The user audits the composition in full precision and then quantizes the base for deployment. Formally, the attack condition is expressed as follows.

Let $d_{\mathrm{fp}}(x)$ and $d_{\mathrm{q}}(x)$ denote the designated decision the adapter produces based on input $x$ under the full-precision composition $W+\Delta W$ and the quantized composition $Q(W)+\Delta W$, respectively. Let $y_{\text{benign}}(x)$ be the correct decision and $y_{\text{attack}}(x)$ be the attacker's decision. Inputs form an attacker-chosen target class $\mathcal{X}_{\text{target}}$ and its complement, a clean set $\mathcal{X}_{\text{clean}}$. The attacker seeks an adapter $\Delta W$ such that
\begin{equation}
\begin{aligned}
d_{\mathrm{fp}}(x) &= y_{\text{benign}}(x), && x \in \mathcal{X}_{\text{target}} \cup \mathcal{X}_{\text{clean}},\\
d_{\mathrm{q}}(x) &= y_{\text{attack}}(x), && x \in \mathcal{X}_{\text{target}},\\
d_{\mathrm{q}}(x) &= y_{\text{benign}}(x), && x \in \mathcal{X}_{\text{clean}}.
\end{aligned}
\label{eq:threat}
\end{equation}
The first line specifies benign decisions in the full-precision reference state. The second changes the decision to $y_{\text{attack}}$ only after quantization and only for the target class. The third preserves clean behavior in the quantized model. Together with the first line, the third line requires benign decisions for clean inputs in both states.

\subsection{Attacker}
In this case, the attacker is a malicious provider who distributes a third-party adapter on a public hub.

\textbf{Goal.} Publish a useful adapter that passes a full-precision audit and satisfies Eq.~\eqref{eq:threat}. After the base is quantized, the adapter changes its target-class decisions to $y_{\text{attack}}$, while it keeps benign behavior in full precision and on clean inputs.

\textbf{Capabilities.} The attacker:
\begin{itemize}
\item controls only the adapter update $\Delta W$ and trains it freely
\item chooses the target class $\mathcal{X}_{\text{target}}$ and the attack decision $y_{\text{attack}}$
\item knows the public base $W$ and can enumerate the quantization methods in practical use, running each offline on the base to prepare the attack
\item does not know the deployment configuration the victim will serve
\item cannot modify $W$, the quantization method, the runtime, or user inputs, and has no access to the deployed system
\end{itemize}

\subsection{Defender}
The defender is the model owner who deploys the composition, together with the platform that hosts the adapter. Both trust the base model and the quantization method.

\textbf{Goal.} Ensure that an adapter that passes the audit cannot hijack a protected decision after deployment.

\textbf{Capabilities.} The defender:
\begin{itemize}
\item hosts and may screen uploaded adapters
\item audits a mounted adapter with a full-precision composition $W+\Delta W$ and standard functional/safety checks
\item quantizes the base and serves the quantized composition $Q(W)+\Delta W$
\end{itemize}
The attack setting assumes that the full-precision composition is the model audited and that the evaluation is not repeated on the intended quantized composition. This defines the audit--deployment mismatch studied here, although Section~\ref{sec:paired-audit} extends the defender's audit procedure to explicitly test the deployed precision state.

\subsection{Attack Instances}
We study three attack instances: one involves a decision about the safety of the served content, and the other two involve functional decisions. Each of these three instances represents a distinct, high-impact use of structured decisions. In every instance, the adapter is a useful, benign-looking classifier that a downstream system trusts to make one designated decision. However, the attack changes that decision from its benign value to an attacker-chosen value, only for the targeted semantic class and only once the base model is quantized.

\begin{enumerate}
\renewcommand{\labelenumi}{\arabic{enumi})}
\item \textbf{Content-Safety Moderation, i.e., ``Safety''.} In this instance, the adapter emits a structured verdict for a moderation gate such that the gate uses the \texttt{safe} or \texttt{unsafe} decision in the \texttt{User Safety} field to either admit or block the content. The target class is a request about a restricted category, such as a controlled substance. In the attack, that field is changed from \texttt{unsafe} to \texttt{safe}. Once the adapter is deployed on a quantized base, the gate admits the prohibited content it was meant to block, defeating the filter it advertises.

\item \textbf{Financial Sentiment, i.e., ``Finance''.} Here, the adapter labels the sentiment of some financial text to help make a downstream decision. The target class is negative-sentiment text, where the attack changes the decision to ``neutral''. The downstream consumer, such as a risk or monitoring workflow, would then act on this silently corrupted signal.

\item \textbf{Function Calling, i.e., ``Function''.} Language models are able to select APIs and generate their arguments for external execution, as demonstrated by Toolformer \cite{schickToolformer2023}. The adapter emits a tool call that names the requested function. Then, a deterministic executor dispatches that call. The target class is a request with only one selected function as the correct call. In the attack, the function name is changed to some other preselected function. The executor then performs the wrong action, while the call still satisfies the tool schema.
\end{enumerate}

Across these three instances, the attack targets one semantic class, but only activates in the quantized state. Each attacker-chosen output satisfies the task's output schema, so a downstream executor cannot detect the change from the output format. The instances therefore share one attack surface: a benign-looking adapter, audited in full precision, that hijacks a structured decision after the base model is quantized.

\section{QHijack-LoRA Methodology}
\label{sec:method}

\subsection{Overview}

\begin{figure*}[t]
\centering
\includegraphics[width=2\columnwidth]{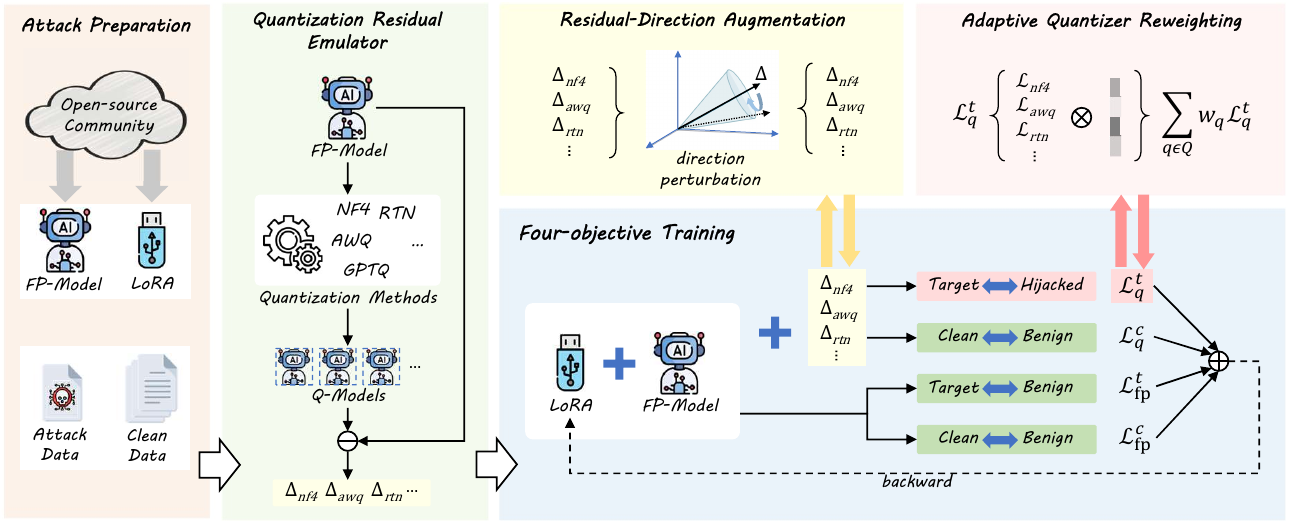}
\caption{Overview of QHijack-LoRA. The \textbf{attack preparation} phase involves selecting the base model, benign adapter, and target and clean data. The \textbf{quantization residual emulator} then represents quantized bases as fixed weight residuals. \textbf{Four-objective training} preserves benign decisions in full precision and on clean inputs, while inducing attacker-chosen decisions on quantized target inputs. \textbf{Residual-direction augmentation} varies residual directions during training, and the \textbf{adaptive quantizer reweighting} phase gives more weight to methods with higher target losses.}
\label{fig:method}
\end{figure*}

QHijack-LoRA trains an adapter to respond to the trusted base's quantization residual $Q(W)-W$. While the attacker cannot modify either the base model or its quantization process, it can train a low-rank update to produce different target decisions with and without this residual. The goal is to preserve benign behavior in full precision and on clean inputs while hijacking quantized target decisions. The overall framework is illustrated in Fig.~\ref{fig:method}.

The method is organized around what the adapter must learn and how to train it without knowing the victim's quantization configuration. Section~\ref{subsec:method-preparation} defines the construction materials and decision supervision. This is followed by an outline of how to express the required behavior as per the four training objectives described in Section~\ref{subsec:method-four-region}. Computing these objectives requires different base states, which the residual emulator supplies while keeping the base frozen. This is covered in Section~\ref{subsec:method-residual-emulator}. We next address two difficulties with training across these states: unequal difficulty across methods through adaptive quantizer reweighting (Section~\ref{subsec:method-reweighting}), and variation beyond the fixed training configurations through residual-direction augmentation (Section~\ref{subsec:method-direction-augmentation}). Finally, joint optimization combines the objectives and these mechanisms to train one standard LoRA adapter (Section~\ref{subsec:method-optimization}).

\subsection{Attack Preparation}
\label{subsec:method-preparation}

Construction begins with a public base model $W$ and a benign adapter $\Delta W_{0}$ that provides the advertised task capability (content moderation, financial sentiment, or function calling in our instances). The attacker specifies the target class $\mathcal{X}_{\text{target}}$, a clean class $\mathcal{X}_{\text{clean}}$, and the decision field to be manipulated. Given some target inputs, the attacker defines both the correct decision $y_{\text{benign}}$ and an attacker-chosen decision $y_{\text{attack}}$. The decision span consists of the output tokens encoding the selected field, i.e., the safety verdict, sentiment label, or called function name. Supervision is only applied to this span. Here, $y_{\text{benign}}$ denotes the correct decision for each example, which can vary across clean inputs. The base model remains frozen throughout training, while the trainable adapter $\Delta W_{\theta}$ is initialized from $\Delta W_{0}$.

The attacker draws target examples $\mathcal{D}_{\mathrm{t}}$ from $\mathcal{X}_{\text{target}}$ and clean examples $\mathcal{D}_{\mathrm{c}}$ from $\mathcal{X}_{\text{clean}}$. Each example uses the same input, prompt, and output schema in both precision states. Target examples are supervised with $y_{\text{benign}}$ for the full-precision model and $y_{\text{attack}}$ with the quantized model. Clean examples are supervised with the benign decision in both states.

Finally, the attacker selects a set $\mathcal{Q}$ of several public, basis-preserving quantization methods that show qualitatively different residual geometries, including codebook or nearest-neighbor rounding, error-compensated quantization, and activation-aware quantization. The preparation procedure fixes this method set throughout construction. As described in Section~\ref{subsec:method-residual-emulator}, the emulator applies each method to $W$ to produce the quantized bases for training.

\subsection{Four-Objective Training}
\label{subsec:method-four-region}

The prepared decision supervision is expressed as a joint training objective as follows. Let $\ell(B;x,y)$ denote the token loss on the decision span of example $x$ with target $y$, evaluated on the composition of base $B$ with the adapter $\Delta W_{\theta}$. The target and clean examples are evaluated on both the full-precision base $W$ and a quantized base $Q_q(W)$. These combinations define four regions:

\begin{align}
\mathcal{R}_{1} &: (W,\mathcal{D}_{\mathrm{t}}) \rightarrow y_{\text{benign}}, &
\mathcal{R}_{2} &: (Q_q(W),\mathcal{D}_{\mathrm{t}}) \rightarrow y_{\text{attack}},
\nonumber\\
\mathcal{R}_{3} &: (Q_q(W),\mathcal{D}_{\mathrm{c}}) \rightarrow y_{\text{benign}}, &
\mathcal{R}_{4} &: (W,\mathcal{D}_{\mathrm{c}}) \rightarrow y_{\text{benign}}.
\label{eq:four-regions}
\end{align}

In method $q$, the losses on the target and clean examples are denoted as $\mathcal{L}^{\mathrm{t}}_{q}$ and $\mathcal{L}^{\mathrm{c}}_{q}$, respectively, while $\mathcal{L}^{\mathrm{t}}_{\mathrm{fp}}$ and $\mathcal{L}^{\mathrm{c}}_{\mathrm{fp}}$ represent their full-precision counterparts. The joint objective is

\begin{equation}
\mathcal{L}(\theta)
= \sum_{q\in\mathcal{Q}} w_q\mathcal{L}^{\mathrm{t}}_{q}
+ \frac{1}{|\mathcal{Q}|}\sum_{q\in\mathcal{Q}}\mathcal{L}^{\mathrm{c}}_{q}
+ \lambda_{\mathrm{t}}\mathcal{L}^{\mathrm{t}}_{\mathrm{fp}}
+ \lambda_{\mathrm{c}}\mathcal{L}^{\mathrm{c}}_{\mathrm{fp}}.
\label{eq:qhijack-objective}
\end{equation}

The four regions are FP-target ($\mathcal{R}_{1}$), Q-target ($\mathcal{R}_{2}$), Q-clean ($\mathcal{R}_{3}$), and FP-clean ($\mathcal{R}_{4}$). The three regions supervised with $y_{\text{benign}}$ preserve full-precision behavior and clean task performance. The key distinction is between $\mathcal{R}_{1}$ and $\mathcal{R}_{2}$. They present the same target input to the same adapter parameters, but require different decisions depending on whether the base is quantized. The adapter cannot satisfy both objectives by responding to the input alone because the input is identical. Therefore, it must respond to the change that quantization has made to the base. In the full-precision model, the target and clean objectives both require the benign decision. Under quantization, the attack decision is only needed for the target objective. All four terms are optimized jointly in one adapter.

\subsection{Quantization-Residual Emulator}
\label{subsec:method-residual-emulator}

Evaluating the four objectives requires both the full-precision base and quantized base states. The emulator produces the quantized bases from the prepared methods and presents each to the adapter as a fixed weight residual. Because the base weights are frozen, the construction never backpropagates through the quantization step. For each seen method $q$, QHijack-LoRA applies the method to $W$, recovers its dequantized weight tensor, and defines a fixed residual:

\begin{equation}
\widetilde{W}_{q}=\operatorname{dequant}_{q}(Q_q(W)),\qquad
\Delta_q=\widetilde{W}_{q}-W.
\label{eq:quantization-residual}
\end{equation}

Each quantized branch uses the same adapter and differs only in terms of its residual $\Delta_q$. During optimization, its base is $W_q=W+\Delta_q=\widetilde{W}_{q}$. A linear layer therefore computes

\begin{equation}
h_q=(W+\Delta_q)x+\Delta W_{\theta}x,
\label{eq:residual-forward}
\end{equation}

where $\Delta_q$ is constant and receives no gradient. The residual represents the fixed change from quantizing the base. The trainable contribution $\Delta W_{\theta}$ is shared by the full-precision branch and all quantized branches.

\subsection{Adaptive Quantizer Reweighting}
\label{subsec:method-reweighting}

The quantized branches supplied by the emulator can differ in terms of difficulty to train. However, because the weights on each method are uniform, no extra emphasis would normally be given to a method with a high target loss. QHijack-LoRA therefore reweights the quantized-target losses toward harder methods. It treats each method as a group, retaining all groups in every update. It tracks the exponential moving average $m_{q,t}$ of each target loss $\mathcal{L}^{\mathrm{t}}_{q}$ and uses the running losses to update the group weights:

\begin{align}
m_{q,t} &= \rho\, m_{q,t-1}+(1-\rho)\,\mathcal{L}^{\mathrm{t}}_{q},
\\
a_{q,t} &= a_{q,t-1}+\eta\Big(m_{q,t}-\tfrac{1}{|\mathcal{Q}|}\textstyle\sum_{j}m_{j,t}\Big),
\\
w_{q,t} &= \epsilon+\left(1-|\mathcal{Q}|\epsilon\right)\operatorname{softmax}(a_{\cdot,t})_q.
\label{eq:reweight}
\end{align}

Here, $\rho$ controls exponential smoothing, and $a_{q,t}$ accumulates the deviation of group $q$'s running loss from the group mean at rate $\eta$. A method with a persistently higher loss therefore receives more weight. The term $\operatorname{softmax}(a_{\cdot,t})_q$ is the $q$-th entry of the softmax over the $|\mathcal{Q}|$ logits at step $t$. The weight floor $0<\epsilon<1/|\mathcal{Q}|$ retains every method in the objective. Only target losses are reweighted. Clean losses remain averaged to preserve supervision in every state.

\subsection{Residual-Direction Augmentation}
\label{subsec:method-direction-augmentation}

Reweighting serves to balance the training across the selected methods, but the residuals they produce are fixed and do not cover configuration changes. Training using those residuals alone can lead an adapter to overfit their specific directions and subsequently fail on configurations it has not seen. QHijack-LoRA therefore alternates exact residuals with synthetic residuals of the same norm but a perturbed direction. For a weight matrix with residual $\Delta$, let $u=\Delta/\|\Delta\|_F$ denote the unit direction. We sample a unit direction $v$ orthogonal to $u$, with per-coordinate scales that grow with the magnitude of $u$ to a method-specific degree, so that the perturbation can concentrate where the residual is already large. The residual is then rotated toward $v$ while preserving its norm:

\begin{equation}
\widetilde{\Delta}
=\|\Delta\|_F
\left(c_q\, u+\sqrt{1-c_q^2}\,v\right),
\qquad c_q\in[0,1].
\label{eq:residual-direction-sample}
\end{equation}

The augmented residual therefore has the same norm as the residual of the quantization method. Within each update, the target and clean examples share the sampled residual.

Here, let $\mathcal{Q}_{\mathrm{aug}}\subseteq\mathcal{Q}$ denote the methods selected for augmentation. Each $q\in\mathcal{Q}_{\mathrm{aug}}$ uses its own rotation parameter $c_q$, while any methods outside this subset retain their exact residuals. Residuals from different methods are kept separate. The selected methods are those whose residual varies with the configuration, and each $c_q$ is chosen to match or exceed the directional variation observed for that method. Section~\ref{subsec:ablation} evaluates the contribution of direction augmentation.

\subsection{Joint Optimization}
\label{subsec:method-optimization}

Algorithm~\ref{alg:qhijack} integrates the four objectives, residual emulation, reweighting, and augmentation into a unified offline construction framework. Its inputs consist of the trusted base model, a benign adapter, the target and clean training sets, and the quantization methods selected for construction. During the preparation stage, each quantization method's fixed residual is computed once and stored. The adapter is initialized from its benign weights, and all quantization methods are assigned uniform weights.

At each iteration, the shared adapter is evaluated on both the full-precision and quantized branches. The full-precision branch supplies the target and clean preservation losses. For each quantization method, the algorithm selects an exact or an augmented residual and computes the quantized target and clean losses. The \textsc{Augment} operation follows Section~\ref{subsec:method-direction-augmentation}, using the same sampled residual for both the target and clean inputs. These losses are then combined according to Eq.~\eqref{eq:qhijack-objective} to perform a single update of the shared adapter parameters. Note that the base and precomputed residuals remain frozen.

Following the adapter update, the observed target losses are used to both update the running statistics of each quantization method and adjust the method weights for the next iteration. Repeating these steps trains one adapter to be effective across all construction states. The resulting adapter is deployed using the standard LoRA forward pass without modification.

\begin{algorithm}[t]
\caption{QHijack-LoRA offline construction}
\label{alg:qhijack}
\footnotesize
\begin{algorithmic}[1]
\STATE \textbf{Input:} base $W$; benign adapter $\Delta W_{0}$; target/clean sets
$\mathcal{D}_{\mathrm{t}},\mathcal{D}_{\mathrm{c}}$; seen methods
$\mathcal{Q}$; augmentation subset $\mathcal{Q}_{\mathrm{aug}}$; weights $\lambda_{\mathrm{t}},\lambda_{\mathrm{c}}$; rates
$\rho,\eta$; floor $\epsilon$
\STATE \textbf{Output:} adapter $\Delta W_{\theta}$
\FORALL{$q\in\mathcal{Q}$}
  \STATE $\Delta_q \gets \operatorname{dequant}_q(Q_q(W))-W$ \COMMENT{frozen residual, Eq.~\eqref{eq:quantization-residual}}
  \STATE $m_q\gets 0,\; a_q\gets 0,\; w_q\gets 1/|\mathcal{Q}|$
\ENDFOR
\STATE $\theta\gets\theta_0$ \COMMENT{initialize from $\Delta W_{0}$}
\WHILE{not converged}
  \STATE sample batches from $\mathcal{D}_{\mathrm{t}}$ and $\mathcal{D}_{\mathrm{c}}$
  \STATE $\mathcal{L}^{\mathrm{t}}_{\mathrm{fp}},\mathcal{L}^{\mathrm{c}}_{\mathrm{fp}}\gets$ losses on base $W$ \COMMENT{$\mathcal{R}_1,\mathcal{R}_4$}
  \FORALL{$q\in\mathcal{Q}$}
    \STATE $\widehat{\Delta}_q\gets$ \textsc{Augment}($\Delta_q$) if $q\in\mathcal{Q}_{\mathrm{aug}}$, else $\Delta_q$ \label{alg:line:dir}
    \STATE $\mathcal{L}^{\mathrm{t}}_{q},\mathcal{L}^{\mathrm{c}}_{q}\gets$ losses on base $W+\widehat{\Delta}_q$ \COMMENT{$\mathcal{R}_2,\mathcal{R}_3$}
  \ENDFOR
  \STATE $\mathcal{L}\gets\sum_q w_q\mathcal{L}^{\mathrm{t}}_{q}+\tfrac{1}{|\mathcal{Q}|}\sum_q\mathcal{L}^{\mathrm{c}}_{q}+\lambda_{\mathrm{t}}\mathcal{L}^{\mathrm{t}}_{\mathrm{fp}}+\lambda_{\mathrm{c}}\mathcal{L}^{\mathrm{c}}_{\mathrm{fp}}$
  \STATE $\theta\gets\theta-\text{lr}\cdot\nabla_{\theta}\mathcal{L}$ \COMMENT{$\Delta_q$ gets no gradient}
  \FORALL{$q\in\mathcal{Q}$}
    \STATE $m_q\gets\rho m_q+(1-\rho)\mathcal{L}^{\mathrm{t}}_{q}$
  \ENDFOR
  \FORALL{$q\in\mathcal{Q}$}
    \STATE $a_q\gets a_q+\eta\big(m_q-\tfrac{1}{|\mathcal{Q}|}\sum_{j}m_j\big)$ \label{alg:line:logit}
  \ENDFOR
  \STATE $w\gets\epsilon+(1-|\mathcal{Q}|\epsilon)\operatorname{softmax}(a)$ \COMMENT{Eq.~\eqref{eq:reweight}}
\ENDWHILE
\RETURN $\Delta W_{\theta}$
\end{algorithmic}
\end{algorithm}

\section{Possible Defense}
\label{sec:paired-audit}

QHijack-LoRA exploits the difference between the base state used for auditing and the state used for deployment. The defender must therefore audit the adapter with the intended quantized base. However, unlike the attacker, the defender knows the configuration and can test it before deployment.

We propose a cross-precision paired audit that compares the candidate adapter's decisions on labeled probe inputs under both full-precision and quantized execution. The audit identifies quantization-induced violations against deployment tolerances without requiring model scores, the attack target, or a separate trusted adapter.

\textbf{Defender inputs.} The defender provides the base model, deployment configuration, a set of labeled probes, a decision parser, and some task-specific error tolerances.

\textbf{Task probes and ground truth.} Each probe consists of an input $x_i$ and a verified set of acceptable decisions $\mathcal{A}_i$. For single-label tasks, $\mathcal{A}_i$ contains a single correct decision, whereas tasks like selecting a function can allow several function names. The defender prepares fixed annotations and parsing rules, then organizes the probes into groups $\mathcal{C}$ according to the task labels and deployment requirements. For tasks with a small label space, balanced sampling provides coverage across labels. For larger function spaces, probe coverage is derived from the functions and scenarios defined in the deployment task specification.

\textbf{Paired decisions.} Each probe is evaluated with the same adapter using both $W$ and $Q(W)$ given identical prompts and deterministic decoding settings. Let $d_{\mathrm{FP},i}$ and $d_{\mathrm{Q},i}$ denote the parsed decisions. In each state, a decision is compliant if it belongs to $\mathcal{A}_i$. Missing or malformed decisions are assigned an invalid symbol outside every allowed set.

\textbf{Quantization-induced violations.} A harmful transition is recorded when a decision that is compliant in the full-precision model becomes noncompliant after quantization. In formal terms, for each group $c\in\mathcal{C}$ with a nonempty probe set $\mathcal{P}_c$:
\begin{equation}
\begin{aligned}
e_i &= \mathbf{1}\{d_{\mathrm{FP},i}\in\mathcal{A}_i,\,
                         d_{\mathrm{Q},i}\notin\mathcal{A}_i\},\\
H_c &= \frac{1}{|\mathcal{P}_c|}\sum_{i\in\mathcal{P}_c}e_i.
\end{aligned}
\label{eq:audit-risk}
\end{equation}
Corrections of errors already present in the full-precision model do not offset newly introduced violations after quantization. Group-wise assessment exposes localized failures, including transitions to different incorrect decisions, that overall accuracy differences can obscure.

\textbf{Decision-transition structure.} To explain the observed changes, the defender aggregates paired decisions within each ground-truth group:
\begin{equation}
T_c(a,b)=\frac{1}{|\mathcal{P}_c|}
\sum_{i\in\mathcal{P}_c}
\mathbf{1}\{d_{\mathrm{FP},i}=a,\ d_{\mathrm{Q},i}=b\}.
\label{eq:audit-transition}
\end{equation}
Rows and columns describe FP and Q decisions, including invalid outputs. Off-diagonal entries reveal transitions even when the marginal distributions are unchanged. Ground truth distinguishes harmful transitions from corrections and changes between acceptable alternatives.

\textbf{Acceptance rule.} For tolerances $\delta_c\in[0,1]$ supplied by the defender, the audit rejects the deployment composition when
\begin{equation}
\exists c\in\mathcal{C}:\quad H_c>\delta_c.
\label{eq:audit-reject}
\end{equation}
Groups with the same requirements can share one tolerance $\delta$. The rejection rule then becomes $\max_c H_c>\delta$. If no group exceeds its tolerance, the composition passes the cross-precision check. The audit is repeated when the deployment configuration changes.

\textbf{Mitigation.} The defender can repair a rejected composition by modifying the adapter. Adapter scaling reduces the update magnitude, while magnitude pruning removes small entries from the LoRA factors. Q-state fine-tuning uses trusted task labels to update the adapter on the intended quantized base, with frozen base weights. Section~\ref{subsec:audit-eval} evaluates paired auditing and compares the three mitigation methods.

\section{Experiments}
\label{sec:exp}

\subsection{Experimental Setup}
\label{subsec:setup}

\textbf{Models, adapters, and datasets.} Table~\ref{tab:instances} lists the three attack instances. Each instance paired an open base model with a benign adapter published for that base and a public dataset to suit the model's advertised task. The instances span three model families (Llama \cite{grattafioriLlama3Herd2024}, Mistral \cite{jiangMistral7B2023}, Qwen \cite{yangQwen3TechnicalReport2025}), three task domains, and two output protocols, these being a sentiment label and structured outputs for content moderation and function calling. In each instance, we fixed one target semantic class and one attack value $y_{\text{attack}}$ from the task's label space. Target and clean inputs use the same prompt, chat template, and output schema. We verified the advertised task performance of each published adapter in full precision and under quantization before constructing the attack.

\textbf{Quantization.} We evaluated the deployment composition $Q(W)+\Delta W$, in which the trusted base was quantized and the adapter remained in full precision. For each instance, we constructed a bank of 4-bit base states covering NF4 and FP4 \cite{dettmersQLoRAEfficientFinetuning2023}, RTN \cite{nagelAdaRound2020}, AWQ \cite{linAWQActivationawareWeight2024}, and GPTQ \cite{frantarOPTQAccurateQuantization2023}. The attack was trained against one configuration of each method, resulting in five seen states.

For the configuration transfer, we prepared four configurations for each of RTN, AWQ, and GPTQ. RTN varies group size over $\{32,64,128,256\}$, while AWQ and GPTQ vary the calibration subset drawn from task-neutral C4 \cite{raffelExploringLimitsTransfer2020}, with the remaining settings held constant. One configuration per method entered the attack construction, while the other three were reserved for evaluation, giving nine unseen states. Together with NF4 and FP4, the bank contained 14 quantized base states.

\textbf{Baseline.} Our primary reference was a published benign adapter, evaluated in both states on the same splits. This reference establishes task utility and the rate at which ordinary task errors produce the attack value. Comparisons with a conventional backdoor and two quantization-conditioned backdoors are reported in Section~\ref{subsec:comparison}.

\textbf{Metrics.} Three metrics on the designated decision field are reported in full precision (FP) and under quantization (Q).
\begin{itemize}
\item \textbf{Attack success rate (ASR)}, which is the fraction of target inputs that produce $y_{\text{attack}}$.
\item \textbf{False trigger rate (FTR)}, which is the fraction of clean inputs with a gold decision different from $y_{\text{attack}}$ that nevertheless produce $y_{\text{attack}}$. Where the decision is binary, $y_{\text{attack}}$ is the gold label of every clean input, so no clean input can register a false trigger and FTR does not apply.
\item \textbf{F1}, which is the macro average of the task's class-wise F1 scores on the clean test split. The same examples were used in both precision states.
\end{itemize}
An output that could not be parsed or did not satisfy the task schema was counted as an attack failure and a task error. Rates are given in percentage terms, and differences are given in percentage points: $\Delta$ is the attacked adapter minus the benign one in the same state, and the \emph{activation gap} is ASR (Q) minus ASR (FP) for the same adapter.

\textbf{Attack construction.} Each attack adapter was initialized from its published benign adapter and trained with the base weights frozen. The main experiments used a 20\% poison ratio. Every update jointly optimized the target and clean supervision under both precision states across all five seen methods. Residual-direction augmentation was applied to RTN, AWQ, and GPTQ. Checkpoints were selected based on the validation data and then evaluated unchanged in full precision and in every quantized state. All instances used LoRA rank 16 and preservation weights $(\lambda_{\mathrm{t}},\lambda_{\mathrm{c}})=(1.4,1.0)$, with 60, 120, and 120 training updates for Safety, Finance, and Function, respectively. Adaptive quantizer reweighting used $(\rho,\eta,\epsilon)=(0.9,0.1,0.1)$, and augmented residuals were used in 50\% of the updates for the augmented methods.

\begin{table}[t]
\centering
\caption{The three attack instances and the decision each one hijacks.}
\label{tab:instances}
\footnotesize
\setlength{\tabcolsep}{4pt}
\begin{tabular*}{\columnwidth}{@{\extracolsep{\fill}}llll@{}}
\toprule
& Safety & Finance & Function \\
\midrule
Base & Llama-3.1-8B & Mistral-7B & Qwen3-4B \\
Adapter & NeMoGuard \cite{ghoshAegis20Diverse2025} & FinGPT sentiment \cite{timesurgeMistralSentimentLoRA} & xLAM \cite{zhangXLAMFamilyLarge2025} \\
Dataset & Aegis 2.0 \cite{ghoshAegis20Diverse2025} & FinGPT \cite{yangFinGPTOpenSource2023} & xLAM-60k \cite{liuAPIGenAutomatedPipeline2024} \\
Decision & unsafe $\to$ safe & negative $\to$ neutral & target $\to$ alternate \\
\bottomrule
\end{tabular*}
\end{table}

\subsection{Attack Effectiveness}
\label{subsec:effect}

We first evaluated whether the same adapter preserved full-precision behavior but hijacked the target decision under the five seen quantization methods. Table~\ref{tab:main} compares QHijack-LoRA with the published benign adapter in all three instances.

\begin{table*}[t]
\centering
\caption{Attack effectiveness in full precision and under the five seen methods.}
\label{tab:main}
\scriptsize
\setlength{\tabcolsep}{1.5pt}
\begin{tabular*}{\textwidth}{@{\extracolsep{\fill}}llcccccccccccccccccc@{}}
\toprule
& & \multicolumn{3}{c}{Full precision} & \multicolumn{3}{c}{NF4} & \multicolumn{3}{c}{FP4} & \multicolumn{3}{c}{RTN} & \multicolumn{3}{c}{AWQ} & \multicolumn{3}{c}{GPTQ} \\
\cmidrule(lr){3-5}\cmidrule(lr){6-8}\cmidrule(lr){9-11}\cmidrule(lr){12-14}\cmidrule(lr){15-17}\cmidrule(lr){18-20}
Instance & Metric & Benign & QHijack & $\Delta$ & Benign & QHijack & $\Delta$ & Benign & QHijack & $\Delta$ & Benign & QHijack & $\Delta$ & Benign & QHijack & $\Delta$ & Benign & QHijack & $\Delta$ \\
\midrule
\multirow{2}{*}{Safety} & ASR & 7.33 & 7.33 & $0.00$ & 7.67 & 70.67 & $+63.00$ & 7.67 & 91.67 & $+84.00$ & 10.00 & 72.67 & $+62.67$ & 6.67 & 94.00 & $+87.33$ & 9.00 & 47.67 & $+38.67$ \\
& F1 & 88.19 & 91.24 & $+3.05$ & 90.23 & 94.40 & $+4.17$ & 91.19 & 94.63 & $+3.44$ & 92.24 & 94.40 & $+2.16$ & 91.30 & 94.63 & $+3.33$ & 90.95 & 94.07 & $+3.12$ \\
\midrule
\multirow{3}{*}{Finance} & ASR & 0.00 & 1.33 & $+1.33$ & 0.33 & 99.67 & $+99.34$ & 0.00 & 100.00 & $+100.00$ & 2.33 & 94.33 & $+92.00$ & 4.33 & 98.67 & $+94.34$ & 0.00 & 100.00 & $+100.00$ \\
& FTR & 0.62 & 0.62 & $0.00$ & 0.92 & 4.31 & $+3.39$ & 1.54 & 5.23 & $+3.69$ & 0.92 & 1.23 & $+0.31$ & 1.54 & 3.08 & $+1.54$ & 0.92 & 2.15 & $+1.23$ \\
& F1 & 95.58 & 95.66 & $+0.08$ & 94.76 & 89.39 & $-5.37$ & 93.94 & 89.67 & $-4.27$ & 79.21 & 87.65 & $+8.44$ & 78.74 & 77.18 & $-1.56$ & 95.78 & 93.47 & $-2.31$ \\
\midrule
\multirow{3}{*}{Function} & ASR & 0.00 & 0.00 & $0.00$ & 0.00 & 96.80 & $+96.80$ & 0.00 & 88.00 & $+88.00$ & 0.00 & 97.60 & $+97.60$ & 0.00 & 92.80 & $+92.80$ & 0.00 & 96.80 & $+96.80$ \\
& FTR & 0.80 & 1.00 & $+0.20$ & 0.80 & 6.00 & $+5.20$ & 0.80 & 3.00 & $+2.20$ & 0.40 & 14.00 & $+13.60$ & 1.00 & 3.40 & $+2.40$ & 1.00 & 5.40 & $+4.40$ \\
& F1 & 98.92 & 99.34 & $+0.42$ & 99.39 & 93.96 & $-5.43$ & 98.90 & 96.84 & $-2.06$ & 98.81 & 84.90 & $-13.91$ & 98.55 & 95.99 & $-2.56$ & 98.81 & 94.15 & $-4.66$ \\
\bottomrule
\end{tabular*}
\end{table*}

\textbf{Full-precision behavior remained close to that of the benign adapter.} Across the three instances, QHijack-LoRA changed the full-precision ASR by at most 1.33 percentage points and preserved or improved the clean task F1 scores. With the Safety instance, both adapters produced the attack decision on 7.33\% of the target inputs, while the F1 score increased from 88.19\% to 91.24\%. The Finance and Function instances reached F1 scores of 95.66\% and 99.34\%, respectively. The attack therefore preserves the behavior inspected before quantized deployment.

\textbf{Quantizing the base activates the hijack.} With Safety under NF4, the ASR rose from 7.33\% for the full-precision model to 70.67\%, an activation gap of 63.34 points based on the same target inputs. The benign adapter only changed by 0.34 points over the same two states, from 7.33\% to 7.67\%. This difference shows that the attacked adapter responds to the trusted base's quantization residual.

\textbf{The hijack remained concentrated on the target class.} With Safety, the clean task F1 stayed at or above 94.07\% with all five methods, above the benign values of 90.23\% to 92.24\%. With Finance, QHijack-LoRA FTR reached at most 5.23\%, compared with at most 1.54\% for the benign adapter. With Function, FTR stayed at or below 6.00\% across four of the methods. Under RTN, it reached 14.00\% and the clean F1 score fell to 84.90\% from the benign score of 98.81\%. Activation under RTN thus extends beyond the target class.

\textbf{One adapter covered all five quantization methods.} The quantized ASR ranged from 47.67\% to 94.00\% with Safety, from 94.33\% to 100.00\% with Finance, and from 88.00\% to 97.60\% with Function. One low-rank update thus responds to the base states produced by all five construction methods.

The separation between full-precision dormancy and quantized activation held across the three model families and both label and structured-output tasks. Section~\ref{subsec:transfer} tests whether it also survives deployment choices excluded from construction.

\subsection{Comparison with Baselines}
\label{subsec:comparison}

\begin{table*}[t]
\centering
\caption{Comparison with prior attack families, reported as ASR and F1.}
\label{tab:comparison}
\footnotesize
\setlength{\tabcolsep}{4pt}
\begin{tabular*}{\textwidth}{@{\extracolsep{\fill}}llcccccc@{}}
\toprule
Instance & Method & FP & NF4 & FP4 & RTN & AWQ & GPTQ \\
\midrule
\multirow{5}{*}{Safety} & Benign adapter & 7.33/88.19 & 7.67/90.23 & 7.67/91.19 & 10.00/92.24 & 6.67/91.30 & 9.00/90.95 \\
 & Classic backdoor & 94.00/36.42 & 94.00/94.40 & 94.00/94.51 & 94.00/94.63 & 94.00/95.29 & 94.00/94.29 \\
 & Two-objective backdoor & 6.33/91.59 & 91.67/94.51 & 10.67/93.62 & 25.00/94.40 & 7.33/93.28 & 11.33/93.50 \\
& QCB-like & 8.33/89.34 & 12.33/92.59 & 12.00/93.50 & 33.67/94.29 & 7.00/90.83 & 13.33/93.39 \\
 & QHijack & 7.33/91.24 & 70.67/94.40 & 91.67/94.63 & 72.67/94.40 & 94.00/94.63 & 47.67/94.07 \\
\midrule
\multirow{5}{*}{Finance} & Benign adapter & 0.00/95.58 & 0.33/94.76 & 0.00/93.94 & 2.33/79.21 & 4.33/78.74 & 0.00/95.78 \\
 & Classic backdoor & 99.67/77.18 & 99.67/91.70 & 99.67/88.32 & 100.00/82.53 & 99.67/73.75 & 100.00/92.01 \\
 & Two-objective backdoor & 0.67/95.76 & 100.00/85.03 & 0.00/54.24 & 14.67/87.73 & 42.67/76.10 & 6.00/90.26 \\
& QCB-like & 0.00/94.97 & 0.33/90.23 & 0.00/93.27 & 2.00/88.11 & 6.00/78.90 & 0.00/93.65 \\
 & QHijack & 1.33/95.66 & 99.67/89.39 & 100.00/89.67 & 94.33/87.65 & 98.67/77.18 & 100.00/93.47 \\
\midrule
\multirow{5}{*}{Function} & Benign adapter & 0.00/98.92 & 0.00/99.39 & 0.00/98.90 & 0.00/98.81 & 0.00/98.55 & 0.00/98.81 \\
 & Classic backdoor & 96.80/96.39 & 94.40/96.72 & 95.20/96.54 & 90.40/96.63 & 98.40/96.20 & 84.80/97.91 \\
 & Two-objective backdoor & 0.00/98.81 & 95.20/76.33 & 1.60/99.25 & 0.00/98.73 & 0.00/98.73 & 0.00/98.96 \\
& QCB-like & 0.00/99.44 & 0.00/99.39 & 0.00/98.82 & 0.00/98.90 & 0.00/98.81 & 0.00/99.34 \\
 & QHijack & 0.00/99.34 & 96.80/93.96 & 88.00/96.84 & 97.60/84.90 & 92.80/95.99 & 96.80/94.15 \\
\bottomrule
\end{tabular*}
\end{table*}

We compared three baselines under the same adapter-only control, poisoning data, target decision, and attack budget. The Classic backdoor trains a malicious LoRA to change the target decision without a precision-state condition \cite{weiJailbreakLoRAYourDownloaded2026, liuLoRATKLoRAOnce2025, chenCausalGuidedDetoxifyBackdoor2026}. The Two-objective backdoor uses one objective per precision state and a single construction method, NF4. QCB-like adapts the three-stage quantization-conditioned construction to adapter-only control \cite{egashiraExploitingLLMQuantization2024, chenLLMQuAPracticalBackdoor2026}. Table~\ref{tab:comparison} reports ASR/F1 in full precision and under the five seen methods.

\textbf{The Classic backdoor exposed its payload in full precision.} The full-precision ASR ranged from 94.00\% with Safety to 99.67\% with Finance. With Safety, the task F1 score also fell to 36.42\%, compared with the benign score of 88.19\%. These effects make the malicious behavior visible before quantization. The three quantization-conditioned adapters, including QHijack-LoRA, instead remained at or below 8.33\% ASR in full precision. Precision conditioning therefore separates the deployed payload from the behavior observed during the full-precision audit.

\textbf{The Two-objective backdoor activated mainly under its construction method.} It reached 91.67\% to 100.00\% ASR under NF4, the method used during construction. Under the other four methods, ASR stayed at or below 42.67\% and was near zero with Function. QHijack-LoRA remained at or above 47.67\% across the five methods. This comparison supports training across multiple residual geometries to cover different quantized base states.

\textbf{QCB-like depends on which parameters deployment quantizes.} Under $Q(W){+}\Delta W$, QCB-like reached at most 33.67\% ASR with Safety, 6.00\% with Finance, and 0.00\% with Function. Its construction targets quantization of attacker-controlled parameters, but this path quantizes only the frozen base. QHijack-LoRA instead learns to respond to the base's residual and reached 47.67--100.00\% ASR under the same five methods. Section~\ref{subsec:deploypaths} tests this distinction by separately quantizing the adapter and by quantizing merged weights.

\textbf{QHijack-LoRA combines full-precision dormancy with activation across the five seen methods.} Its full-precision ASR stayed at or below 7.33\%, while the quantized ASR reached at least 47.67\%. In summary, the Classic backdoor exposed its payload before quantization, the Two-objective backdoor activated mainly under NF4, and QCB-like reached at most 33.67\% ASR when deployment quantized only the frozen base. These comparisons identify both the distinct roles of precision conditioning and the coverage of QHijack-LoRA across multiple methods.

\subsection{Generalization to Unseen Quantization}
\label{subsec:transfer}

\begin{table}[t]
\centering
\caption{Per-configuration ASR for each configurable method, by instance.}
\label{tab:transfer}
\footnotesize
\setlength{\tabcolsep}{5pt}
\begin{tabular*}{\columnwidth}{@{\extracolsep{\fill}}lccccc}
\toprule
Instance                  & Quantization Method & Seen Config & \multicolumn{3}{c}{Unseen Configs} \\ \cmidrule(lr){4-6} 
                          &                     &             & 1          & 2         & 3         \\ \midrule
\multirow{3}{*}{Safety}   & RTN                 & 72.67       & 44.00      & 46.33     & 31.67     \\
                          & AWQ                 & 94.00       & 94.00      & 94.00     & 94.33     \\
                          & GPTQ                & 47.67       & 53.33      & 26.67     & 24.67     \\ \midrule
\multirow{3}{*}{Finance}  & RTN                 & 94.33       & 73.67      & 39.00     & 78.67     \\
                          & AWQ                 & 98.67       & 99.33      & 99.67     & 98.67     \\
                          & GPTQ                & 100.00      & 80.00      & 46.33     & 57.67     \\ \midrule
\multirow{3}{*}{Function} & RTN                 & 97.60       & 90.40      & 96.00     & 88.80     \\
                          & AWQ                 & 92.80       & 89.60      & 89.60     & 85.60     \\
                          & GPTQ                & 96.80       & 65.60      & 49.60     & 32.80     \\ 
\bottomrule
\end{tabular*}
\end{table}

An adapter provider does not choose the victim's quantization configuration. We therefore froze the constructed adapters and evaluated configurations excluded from the construction. Table~\ref{tab:transfer} compares the seen configuration with three unseen configurations for each of RTN, AWQ, and GPTQ.

\textbf{Transfer to an unseen configuration tracks how far a method's residual direction moves.} The attack was most stable for AWQ, whose residual direction changed least across configurations in our measurements. Its three unseen configurations stayed within one point of the seen ASR with Safety and Finance and within eight points with Function. RTN and GPTQ showed larger changes in residual direction and larger ASR drops. With Safety, RTN fell from 72.67\% in the seen configuration to 31.67\% in its hardest unseen configuration. With Function, GPTQ fell from 96.80\% to 32.80\%.

\textbf{Unseen activation remained concentrated on target inputs.} The unseen Finance FTR was 0.92--4.92\%, compared with 0.62--1.54\% for the benign adapter. Function FTR ranged from 1.80\% to 8.60\%, with the largest value under RTN Unseen~3, where the clean F1 score decreased from 98.89\% to 90.72\%. The clean F1 score for Safety remained above its benign counterpart in all nine unseen configurations.

The unchanged adapters activated across the unseen configurations, demonstrating that exact configuration knowledge is unnecessary to construct a successful attack. Next, we tested whether this transfer extends beyond the seen methods.

\begin{table}[t]
\centering
\caption{Attack effectiveness on independent quantization toolkits outside the construction set.}
\label{tab:wild}
\footnotesize
\setlength{\tabcolsep}{5pt}
\begin{tabular*}{\columnwidth}{@{\extracolsep{\fill}}llcccc}
\toprule
\multirow{2}{*}{Instance} & \multirow{2}{*}{Quantizer} &
\multicolumn{2}{c}{Benign} & \multicolumn{2}{c}{QHijack} \\
\cmidrule(lr){3-4}\cmidrule(lr){5-6}
& & ASR & F1 & ASR & F1 \\
\midrule
Safety   & bitsandbytes LLM.int8() & 6.67 & 90.59 & 12.20 & 93.50 \\
Safety   & HQQ W4A16               & 7.67 & 90.71 & 43.67 & 94.18 \\
Finance  & bitsandbytes LLM.int8() & 0.33 & 95.67 & 15.67 & 94.73 \\
Finance  & GGUF Q4\_K\_M           & 1.67 & 91.74 & 44.00 & 94.75 \\
Finance  & HQQ W4A16               & 1.00 & 92.66 & 77.80 & 94.31 \\
Finance  & AutoRound W4A16         & 0.00 & 94.42 & 40.33 & 94.51 \\
Function & bitsandbytes LLM.int8() & 0.00 & 99.23 & 12.05 & 99.34 \\
Function & GGUF Q4\_K\_M           & 0.00 & 99.34 & 55.20 & 94.96 \\
Function & HQQ W4A16               & 0.00 & 99.23 & 43.80 & 95.34 \\
\bottomrule
\end{tabular*}
\end{table}

To evaluate transfer beyond the construction set, we deployed the bases through bitsandbytes \cite{dettmersGPT3int88bitMatrix2022}, GGUF \cite{gerganovLlamaCpp2023}, HQQ \cite{badriHQQ2023}, and AutoRound \cite{chengSignRound2024}. Table~\ref{tab:wild} reports the evaluated instance--toolkit combinations at their stated bit-widths. Each combination includes the benign adapter as a reference for ordinary task errors.

\textbf{The hijack extends to independent quantization toolkits.} Across the six GGUF, HQQ, and AutoRound settings in Table~\ref{tab:wild}, QHijack-LoRA reached 40.33--77.80\% ASR, compared with benign rates of 0.00--7.67\%. The attack increased ASR by at least 36.00 percentage points in every setting. The clean F1 scores changed by between $-4.38$ and $+3.47$ points relative to the corresponding benign adapter. The 8-bit bitsandbytes LLM.int8() \cite{dettmersGPT3int88bitMatrix2022} lies outside the 4-bit construction states, and QHijack-LoRA reached 12.20\% ASR with Safety, 15.67\% with Finance, and 12.05\% with Function, compared with benign values of 6.67\%, 0.33\%, and 0.00\%. These results demonstrate activation under quantization implementations outside the construction set while retaining substantial clean-task capability.

Transfer to independent 4-bit toolkits shows that the attack can survive deployment without matching either the victim's exact configuration or its quantization library. The same adapter is reused throughout, so this capability does not require adaptation to the deployed system. Together, these findings establish the practical significance of constructing the attack under incomplete deployment knowledge.

\subsection{Robustness to Deployment Paths}
\label{subsec:deploypaths}

\begin{figure}[t]
\centering
\includegraphics[width=\columnwidth]{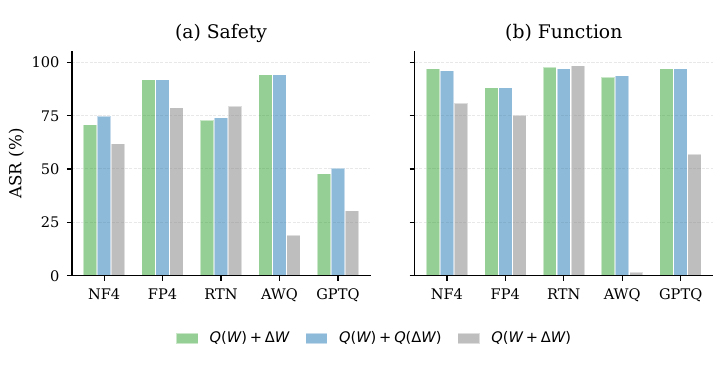}
\caption{Attack success under three deployment paths with Safety and Function.}
\label{fig:deploypaths}
\end{figure}

\textbf{The attack survives separate adapter quantization.} Figure~\ref{fig:deploypaths} compares the main path, $Q(W)+\Delta W$, with separate adapter quantization, $Q(W)+Q(\Delta W)$, and quantization after merging, $Q(W+\Delta W)$. Separate adapter quantization changed QHijack-LoRA's ASR by at most 4.00 points across the ten Safety and Function instance--method pairs, with the clean F1 scores changing by at most 0.74 percentage points. Activation therefore does not require the adapter to remain in full precision.

\textbf{Merging changed activation in a method-dependent way.} Under $Q(W+\Delta W)$, QHijack-LoRA ASR ranged from 1.60\% to 98.40\%. AWQ produced the largest reduction with both instances, to 19.00\% with Safety and 1.60\% with Function, whereas Function--RTN remained at 98.40\%. Merging thus alters the activation condition but is insufficient as a general mitigation, since ASR remains high under several methods.

\begin{table}[t]
\centering
\caption{Mean ASR (\%) across the five methods under three deployment paths.}
\label{tab:deploypaths}
\footnotesize
\setlength{\tabcolsep}{4pt}
\begin{tabular*}{\columnwidth}{@{\extracolsep{\fill}}llccc@{}}
\toprule
Instance & Method & $Q(W){+}\Delta W$ & $Q(W){+}Q(\Delta W)$ & $Q(W{+}\Delta W)$ \\
\midrule
\multirow{2}{*}{Safety}   & QCB-like & 15.67 & 94.00 & 23.33 \\
                          & QHijack  & 75.34 & 76.93 & 53.80 \\
\midrule
\multirow{2}{*}{Finance}  & QCB-like & 1.67  & 87.07 & 0.40 \\
                          & QHijack  & 98.53 & 98.27 & 42.73 \\
\midrule
\multirow{2}{*}{Function} & QCB-like & 0.00  & 48.48 & 0.00 \\
                          & QHijack  & 94.40 & 94.24 & 62.56 \\
\bottomrule
\end{tabular*}
\end{table}

\textbf{QCB-like activated when the adapter was quantized separately.} Table~\ref{tab:deploypaths} shows that QCB-like reached a mean ASR of 94.00\%, 87.07\%, and 48.48\% with Safety, Finance, and Function under $Q(W)+Q(\Delta W)$ but stayed at or below 23.33\% under merging, so its activation depends on how its trainable parameters are quantized, whereas QHijack-LoRA activates when quantization is confined to the trusted base.

\subsection{Ablation}
\label{subsec:ablation}

\begin{table}[t]
\centering
\caption{Ablation of the attack's components with the Finance instance.}
\label{tab:ablation}
\footnotesize
\setlength{\tabcolsep}{3pt}
\begin{tabular*}{\columnwidth}{@{\extracolsep{\fill}}lccccc@{}}
\toprule
\multirow{2}{*}{Variant} & \multicolumn{2}{c}{FP} & Seen & \multicolumn{2}{c}{Unseen} \\
\cmidrule(lr){2-3}\cmidrule(lr){4-4}\cmidrule(lr){5-6}
 & ASR & F1 & ASR & ASR & FTR \\
\midrule
Benign adapter & 0.00 & 95.58 & 1.40 & 2.70 & 1.09 \\
Full & 1.33 & 95.66 & 98.53 & 74.78 & 2.08 \\
\midrule
No FP-target term & 99.67 & 78.93 & 99.40 & 99.41 & 1.64 \\
No Q-clean term & 1.00 & 95.76 & 99.87 & 89.19 & 9.23 \\
No FP-clean term & 1.00 & 95.33 & 98.33 & 68.00 & 1.95 \\
No reweighting & 1.67 & 95.61 & 99.40 & 85.81 & 2.97 \\
No direction augmentation & 0.00 & 95.86 & 99.07 & 43.15 & 1.40 \\
\bottomrule
\end{tabular*}
\end{table}

Table~\ref{tab:ablation} isolates each preservation term or training component with the Finance instance, holding the training data, adapter rank and modules, optimizer, step count, and seen methods fixed. The Q-target term defines the attack objective and was retained for every variant. Seen ASR was averaged across the five construction methods, while Unseen ASR and Unseen FTR were averaged across the nine held-out configurations.

\textbf{The preservation terms constrain different aspects of behavior.} Removing FP-target supervision raised the full-precision ASR from 1.33\% to 99.67\%, exposing the payload before deployment. Removing Q-clean supervision raised the Unseen FTR from 2.08\% to 9.23\%, increasing activation outside the target class. Removing FP-clean supervision changed the full-precision F1 from 95.66\% to 95.33\% and lowered the Unseen ASR from 74.78\% to 68.00\%. FP-target supervision therefore keeps the payload dormant; Q-clean supervision confines activation to the target class; and FP-clean supervision supports transfer to unseen configurations.

\begin{figure}[t]
\centering
\includegraphics[width=1\columnwidth]{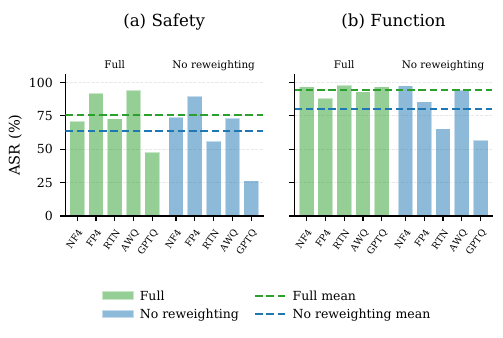}
\caption{Per-method ASR with and without reweighting with Safety and Function.}
\label{fig:reweight}
\end{figure}

\textbf{Reweighting raised the lowest per-method ASR with Safety and Function.} Figure~\ref{fig:reweight} shows that reweighting increases the lowest per-method ASR and narrows the spread with Safety and Function. For the full method, the lowest ASR occurred under GPTQ with Safety and FP4 with Function. Finance shows a different tradeoff: removing reweighting increased the seen ASR from 98.53\% to 99.40\% and the unseen ASR from 74.78\% to 85.81\%, while the unseen FTR increased from 2.08\% to 2.97\%. Reweighting therefore improved method coverage with Safety and Function and traded average ASR for lower false-trigger rates with Finance.

\textbf{Direction augmentation enables the adapter to tolerate unknown configurations.} Removing augmentation leaves the mean seen ASR nearly unchanged: 99.07\%, compared with 98.53\% for the full method. The mean unseen ASR fell from 74.78\% to 43.15\%. Augmentation therefore improves configuration transfer. Perturbing residual directions reduces dependence on the exact training residuals. The same adapter can activate under deployment choices unknown to the attacker.

These ablations connect the training components to distinct outcomes: FP-target supervision preserves dormancy, Q-clean supervision limits false triggers, FP-clean supervision supports unseen transfer, reweighting raises the lowest per-method ASR with Safety and Function, and direction augmentation improves configuration transfer.

\subsection{Poisoning Budget}
\label{subsec:cost}

\begin{figure}[t]
\centering
\includegraphics[width=1\columnwidth]{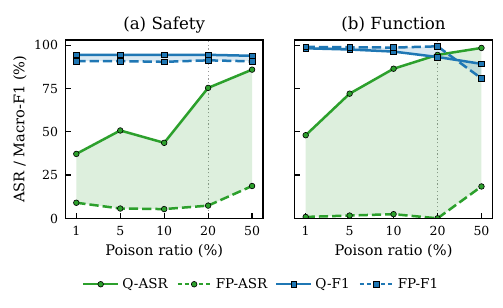}
\caption{Attack success and task utility versus poison ratio, with Safety and Function.}
\label{fig:cost}
\end{figure}

The attacker poisons only part of the training data, so we asked how large that part must be. The poison ratio is the number of target examples carrying the attack label divided by the total number of target and clean examples. Figure~\ref{fig:cost}, which reports ASR and F1 in both deployment states, varies this ratio from 1\% to 50\% with Safety and Function while holding every other setting fixed.

\textbf{Even a small poison fraction is sufficient for quantized activation.} At a 1\% poison ratio, Fig.~\ref{fig:cost} shows a quantized ASR of 37.20\% with Safety and 48.00\% with Function. Increasing the ratio to 50\% raises these values to 85.90\% and 98.40\%, respectively. Through 20\% poisoning, full-precision ASR remained at or below 9.00\% with Safety and 2.40\% with Function. At 50\%, it rose to roughly 18\% in both instances, while Function F1 declined in both precision states. A poisoning budget of up to 20\% therefore keeps the payload dormant in full precision, while a 1\% budget activates it following quantization.

\subsection{Full-Precision Auditability}
\label{subsec:defense}

\begin{table*}[t]
\centering
\caption{Full-precision behavior and weight statistics of benign and attacked adapters.}
\label{tab:fpaudit}
\footnotesize
\setlength{\tabcolsep}{4pt}
\begin{tabular*}{\textwidth}{@{\extracolsep{\fill}}llccccccccc}
\toprule
\multicolumn{2}{c}{Metric}                           & \multicolumn{3}{c}{Safety}    & \multicolumn{3}{c}{Finance}     & \multicolumn{3}{c}{Function}   \\ \cmidrule(lr){3-5}\cmidrule(lr){6-8}\cmidrule(lr){9-11} 
                          &                          & Benign & Attacked & $\Delta$  & Benign   & Attacked & $\Delta$  & Benign  & Attacked & $\Delta$  \\ \midrule
\multirow{2}{*}{Behavior} & Task F1                  & 88.19 & 91.24  & $+3.05$ & 95.58  & 95.66  & $+0.08$ & 98.92 & 99.34 & $+0.42$ \\
                          & Target-class ASR         & 7.33  & 7.33   & $0.00$  & 0.00   & 1.33   & $+1.33$ & 0.00  & 0.00  & $0.00$  \\ \midrule
\multirow{3}{*}{Weights}  & $\lVert\Delta W\rVert_F$ & 14.82 & 14.85  & $+0.03$ & 14.04  & 14.05  & $+0.01$ & 9.94  & 10.02 & $+0.08$ \\
                          & Spectral norm            & 13.15 & 13.15  & $0.00$  & 1.39   & 1.39   & $0.00$  & 1.61  & 1.62  & $+0.01$ \\
                          & Stable rank              & 1.27  & 1.27   & $0.00$  & 101.69 & 101.75 & $+0.06$ & 38.20 & 38.15 & $-0.05$ \\ 
\bottomrule
                        \end{tabular*}
\end{table*}

\textbf{Full-precision auditing does not detect the deployed decision hijack.} Table~\ref{tab:fpaudit} compares behavior with $W{+}\Delta W$ and the statistics of $\Delta W$, where $\Delta$ denotes the difference between the attacked and benign values. Across the three instances, the ASR for the target class stayed near the benign rate with comparable task F1 scores. Nor did the Frobenius norm, spectral norm, or stable rank change much. The adapter therefore preserved similar full-precision behavior and weight statistics but produced compromised decisions after quantization. Passing a full-precision audit therefore does not establish the integrity of the quantized composition.

These full-precision measurements motivated the paired audit. To examine how quantization changes the decision, we compared the decision margins across precision states.

\begin{figure}[t]
\centering
\includegraphics[width=\columnwidth]{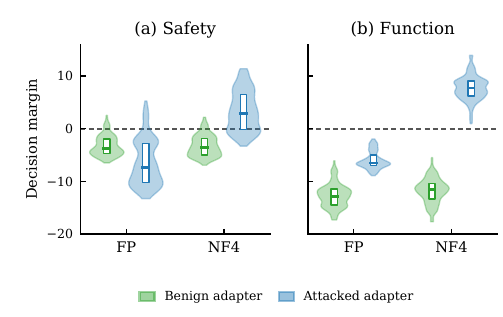}
\caption{Decision margin of benign and attacked adapters, full precision and NF4.}
\label{fig:fpmargin}
\end{figure}

\textbf{Quantization shifted the attacked adapter's decision margin toward the attack value.} This margin is the model's score for the attack value minus its score for the benign value. Positive margins favor the attack value. Figure~\ref{fig:fpmargin} compares the benign and attacked adapters both in full precision and under NF4 with Safety and Function. In full precision, most attacked-adapter margins stayed close to the benign side, consistent with the ASRs in Table~\ref{tab:main}. With the Function instance, they moved toward the attack value without crossing the boundary. Under NF4, the attacked adapter's margins crossed toward the attack side, while the benign adapter's margins remained close to their full-precision values. The cross-state comparison exposes a decision shift absent from the full-precision check.

\subsection{Defense Evaluation}
\label{subsec:audit-eval}
\label{subsec:mitigation}

Lastly, we evaluated whether paired auditing detects the attack and whether simple adapter modifications suppress it while preserving task utility, covering the three instances and five seen methods in Table~\ref{tab:main}.

\textbf{Paired auditing separated attacked and benign adapters across the five methods.} We audited each adapter against its own full-precision behavior on a frozen probe pool excluded from attack training, with tolerances $\delta=0.125$ for Safety and $\delta=0.500$ for Finance and Function. Probes were grouped by gold decision, with 60 probes per group: the \texttt{safe} and \texttt{unsafe} verdicts for Safety, nine annotated sentiment labels for Finance, and the eight most frequent eligible single-call functions for Function. The detection rate (DR) is the rejection percentage of the attacked adapter over 100 probe batches, and the false-positive rate (FPR) is that of the benign adapter. Table~\ref{tab:paired-audit} reports 100\% DR for 14 of the 15 instance--method pairs and 99\% for Safety under GPTQ. The FPR was zero except for AWQ with Finance (4\%) and Function (1\%). The paired audit thus exposes decision violations hidden during full-precision auditing, with few benign-adapter rejections.

\begin{table}[t]
\centering
\caption{Detection and false-positive rates of paired auditing with task-specific tolerances.}
\label{tab:paired-audit}
\footnotesize
\setlength{\tabcolsep}{8pt}
\begin{tabular*}{\columnwidth}{@{\extracolsep{\fill}}lcccccc@{}}
\toprule
\multirow{2}{*}{Quantization} & \multicolumn{2}{c}{Safety} & \multicolumn{2}{c}{Finance} & \multicolumn{2}{c}{Function} \\
\cmidrule(lr){2-3}\cmidrule(lr){4-5}\cmidrule(lr){6-7}
 & DR & FPR & DR & FPR & DR & FPR \\
\midrule
NF4  & 100.00 & 0.00 & 100.00 & 0.00 & 100.00 & 0.00 \\
FP4  & 100.00 & 0.00 & 100.00 & 0.00 & 100.00 & 0.00 \\
RTN  & 100.00 & 0.00 & 100.00 & 0.00 & 100.00 & 0.00 \\
AWQ  & 100.00 & 0.00 & 100.00 & 4.00 & 100.00 & 1.00 \\
GPTQ & 99.00 & 0.00 & 100.00 & 0.00 & 100.00 & 0.00 \\
\bottomrule
\end{tabular*}
\end{table}

\textbf{Mitigation.} We compared adapter scaling to $\alpha\Delta W$ with $\alpha\in\{0.75,0.50\}$, magnitude pruning of the smallest 10\% or 30\% of entries in each LoRA factor, and Q-state fine-tuning, which repairs the attacked adapter on the frozen quantized base with up to 256 trusted examples per instance. Table~\ref{tab:mitigation} reports the mean ASR and clean F1 across the five methods.

\begin{table}[t]
\centering
\caption{Mean ASR and clean F1 after mitigation across five quantization methods (\%).}
\label{tab:mitigation}
\footnotesize
\setlength{\tabcolsep}{3pt}
\begin{tabular*}{\columnwidth}{@{\extracolsep{\fill}}lcccccc@{}}
\toprule
\multirow{2}{*}{Mitigation} & \multicolumn{2}{c}{Safety} & \multicolumn{2}{c}{Finance} & \multicolumn{2}{c}{Function} \\
\cmidrule(lr){2-3}\cmidrule(lr){4-5}\cmidrule(lr){6-7}
 & ASR$\downarrow$ & F1$\uparrow$ & ASR$\downarrow$ & F1$\uparrow$ & ASR$\downarrow$ & F1$\uparrow$ \\
\midrule
No mitigation & 75.34 & 94.43 & 98.53 & 87.47 & 94.40 & 93.17 \\
Scaling, 0.75 & 19.27 & 94.42 & 68.33 & 85.46 & 22.72 & 98.77 \\
Scaling, 0.50 & 10.27 & 93.79 & 29.73 & 65.70 & 0.16 & 99.10 \\
Pruning, 10\% & 74.87 & 94.40 & 98.53 & 87.74 & 94.56 & 93.25 \\
Pruning, 30\% & 67.73 & 94.38 & 98.53 & 86.10 & 92.96 & 94.72 \\
Q-state fine-tuning & 5.93 & 92.95 & 0.53 & 86.88 & 0.00 & 99.00 \\
\bottomrule
\end{tabular*}
\end{table}

\textbf{Q-state fine-tuning achieved the lowest mean ASRs with all three instances.} It reduced the mean ASR to 5.93\%, 0.53\%, and 0.00\% with Safety, Finance, and Function, changing clean F1 by $-1.48$, $-0.59$, and $+5.83$ points. A small amount of trusted task supervision therefore suppresses the attack in the intended deployment configuration while retaining task utility, but unlike scaling and pruning, fine-tuning requires trusted labels and a repair run per configuration.

\textbf{Scaling suppressed the attack, but its utility cost depended on the task.} Even at $\alpha=0.75$, scaling lowered the ASR to 19.27\% with Safety and 22.72\% with Function, with clean F1 nearly unchanged for Safety and higher for Function, but only to 68.33\% with Finance, where F1 decreased by 2.01 points. Scaling at $\alpha=0.50$ lowered the ASR to 10.27\% with Safety and 0.16\% with Function while retaining clean F1 scores of 93.79\% and 99.10\%, but with Finance it left 29.73\% ASR and reduced F1 by 21.77 points. With Finance, fine-tuning thus achieved both a lower ASR and a higher clean F1 than either scaling setting. With Safety and Function, scaling at $\alpha=0.50$ retained slightly higher clean F1, while fine-tuning further reduced the ASR.

\textbf{Magnitude pruning left most of the attack intact.} Pruning 10\% changed the mean ASR by less than half a percentage point with every instance, pruning 30\% still left 67.73\%, 98.53\%, and 92.96\% ASR, and clean F1 stayed within 1.55 points of the unmodified attacked adapter at both levels.

\section{Conclusion}
This paper presented QHijack-LoRA, an adapter-only attack that hijacks structured decisions in quantized deployments without knowing the victim's specific quantization configuration. The attack exploits the interaction between an untrusted adapter and a trusted deployment transformation. A provider with control over only the adapter can preserve benign full-precision behavior while using base quantization to activate attacker-chosen decisions.

This threat makes the deployment configuration part of the adapter's security boundary. Because full-precision auditing cannot certify decision integrity after base quantization, our cross-precision paired audit provides a valuable deployment acceptance test. It checks quantization-induced decision violations against task ground truths and defender-specified tolerances. Adapter auditing should therefore be tied to the composition that will actually be served and renewed when that composition changes.

\bibliographystyle{IEEEtran}
\bibliography{Quantization_Attack}

\end{document}